\documentclass{ws-ijmpa}
\usepackage[super,compress]{cite}
\usepackage{graphicx}
\usepackage{slashed}
\usepackage{textcomp}
\begin{document}
\markboth{Authors' Names}
{Instructions for Typing Manuscripts (Paper's Title)}

%
\catchline{}{}{}{}{}
%

\title{Decays $Z\to \gamma \gamma$ and $Z\to gg$ in the Standard Model Extension}

\author{J. CASTRO--MEDINA}

\address{Facultad de Ciencias F\'{\i}sico Matem\'aticas,
Benem\'erita Universidad Aut\'onoma de Puebla, Apartado Postal
1152, Puebla, Puebla, M\'exico.}

\author{H. NOVALES--SANCHEZ}

\address{Facultad de Ciencias F\'{\i}sico Matem\'aticas,
Benem\'erita Universidad Aut\'onoma de Puebla, Apartado Postal
1152, Puebla, Puebla, M\'exico. \\ hnovales@fcfm.buap.mx}

\author{J. J. TOSCANO}

\address{Facultad de Ciencias F\'{\i}sico Matem\'aticas,
Benem\'erita Universidad Aut\'onoma de Puebla, Apartado Postal
1152, Puebla, Puebla, M\'exico,
\\ and Facultad de Ciencias F\'isico Matem\'aticas, Universidad Michoacana de San Nicol\'as de Hidalgo, Avenida Francisco J. M\'ujica S/N, 58060, Morelia, Michoac\'an, M\'exico. \\ jtoscano@fcfm.buap.mx}

\author{E. S. TUTUTI}

\address{Facultad de Ciencias F\'isico Matem\'aticas, Universidad Michoacana de San Nicol\'as de Hidalgo, Avenida Francisco J. M\'ujica S/N, 58060, Morelia, Michoac\'an, M\'exico. \\ tututi@umich.mx}

\maketitle

\begin{history}
\received{Day Month Year}
\revised{Day Month Year}
\end{history}

\begin{abstract}
The $Z\to \gamma \gamma$ and $Z\to gg$ decays are studied in the context of the renormalizable version of the Standard Model Extension.
The $CPT$--odd $\bar{\psi}\gamma_5 \slashed{b}\psi$ bilinear interaction, which involves the constant background field $b_\alpha$ and which has been a subject of interest in literature, is considered.
It is shown that the $Z\to \gamma \gamma$ and $Z\to gg$ decays, which are strictly zero in the standard model, can be generated radiatively at the one-loop level. It is found that these decays are gauge invariant and free of ultraviolet divergences, and that the corresponding decay widths only depend on the spatial component of the background field $b$.
\end{abstract}

\ccode{PACS numbers: 11.30.Cp, 13.38.Dg}


\section{Introduction}
\label{I}

The trilinear $VVV$ $(V=\gamma,Z)$ and quartic $VVVV$
neutral couplings are quite suppressed, as they first arise at the
one--loop level\footnote{The exception is the $ZZZZ$ vertex, which
can be induced at the tree level via the interchange of a neutral
Higgs boson.} in the Standard Model (SM) and in most of its
renormalizable extensions. Since new physics effects could be more
apparent in those processes which are quite suppressed or
forbidden in the SM, these couplings constitute a good mechanism
for investigating possible signals of physics beyond the Fermi
scale. The charged counterparts $WWV$ and $WWVV$, of these neutral couplings, arise at the tree level and are related by gauge invariance in the sense that they are simultaneously induced by $SU_L(2)\times U_Y(1)$--invariant operators~\cite{EL1,EL2}. However, this type of gauge connection does not exist between the $VVV$ and $VVVV$ couplings~\cite{TT1,TT2}. While the $VVVV$ couplings receive one--loop contributions from
both fermionic and bosonic particles\footnote{Light by light scattering has been explored in
diverse contexts in Refs.~\cite{LbyL1,LbyL2,LbyL3,LbyL4,LbyL5,LbyL6,LbyL7,LbyL8,LbyL9,LbyL10,LbyL11,LbyL12,LbyL13}, works about the $Z\gamma \gamma \gamma$ vertex can be found in Refs.~\cite{Z3gamma1,Z3gamma2,Z3gamma3,Z3gamma4}, the $ZZ\gamma \gamma$ vertex is studied in Refs.~\cite{2Z2gamma1,2Z2gamma2,2Z2gamma2y1/2,2Z2gamma3,2Z2gamma4,2Z2gamma5,2Z2gamma6,2Z2gamma7,2Z2gamma8,2Z2gamma9}, and quartic neutral couplings among $Z$ gauge bosons, photons, and gluons have been investigated in Refs.~\cite{vg1,vg2,vg3,vg4,vg5,vg6}}, the $VVV$ ones
are exclusively generated by fermionic triangles~\cite{SMVVV1,SMVVV2,SMVVV3}. Intimately related with the absence of a gauge link between the $VVV$ and
$VVVV$ couplings is the fact that none of the $VVV$ vertices
reflects the non--Abelian nature of the electroweak group, as the
Yang--Mills sector neither induces these couplings at the
tree--level nor contribute to them at the one--loop order. This is in
contrast with the quartic $VVVV$ couplings, which are generated by
the Yang--Mills sector at the one--loop level~\cite{LbyL1,LbyL2,LbyL3,LbyL4,LbyL5,LbyL6,LbyL7,LbyL8,LbyL9,LbyL10,LbyL11,LbyL12,LbyL13,Z3gamma1,Z3gamma2,Z3gamma3,Z3gamma4,2Z2gamma1,2Z2gamma2,2Z2gamma2y1/2,2Z2gamma3,2Z2gamma4,2Z2gamma5,2Z2gamma6,2Z2gamma7,2Z2gamma8,2Z2gamma9,vg1,vg2,vg3,vg4,vg5,vg6}.
The
trilinear $VVV$ couplings are indeed severely restricted by the Bose
and Lorentz symmetries. It turns out that if both Bose symmetry
and Lorentz invariance are simultaneously respected, the $VVV$
coupling vanishes when the three bosons are real, though they can
exist when at least one of the particles is off--shell~\cite{SMVVV1,SMVVV2,SMVVV3}. As a
consequence, the $Z\to \gamma \gamma$ decay is forbidden. This
result is known as the Landau--Yang's theorem~\cite{LYT1,LYT2}, which
establishes that a vector particle cannot decay into two massless
vector particles. This means that the $Z$ decay into two gluons is also
forbidden\footnote{It is important to stress that the Landau--Yang's theorem is actually only applicable to on-shell particles~\cite{SMVVV1,SMVVV2,SMVVV3,Rus}. For instance, a virtual $Z$ gauge boson can contribute to the $gg \to Z \to \gamma \gamma$ reaction~\cite{Moretti}. }. This theorem invokes Bose symmetry and rotational invariance arguments.  Nevertheless, as soon as this requirement is relaxed, the $VVV$ vertex with the
three bosons on--shell can exist. In particular, the rare decays of the $Z$ gauge boson $Z\to \gamma \gamma$ and $Z\to gg$ are allowed.

In this paper, we are interested in studying the decays $Z\to \gamma \gamma$ and $Z\to gg$ in the context of the minimal Lorentz-- and $CPT$--violating Standard Model Extension~\cite{SME1,SME2} (SME), which is a renormalizable extension of the SM that incorporates Lorentz violation and $CPT$ nonconservation in a model independent fashion through the effective Lagrangian technique. While motivated by specific scenarios in the context of string theory~\cite{KS1,KS2}, general relativity with spontaneous symmetry breaking~\cite{QG1,QG2,QG3,QG4}, space-time-varying coupling constants~\cite{KSTV}, nontrivial topology of space~\cite{K}, or field theories formulated in a noncommutative space-time~\cite{Snyder,SWM,NCYM1,NCYM2,NCYM3,NCYM4,NCYM5,NCSM}, the SME is beyond these specific ideas due to its generality, which is the main advantage of effective field theories. Thus the SME provides us with a powerful tool for investigating $CPT$ nonconservation and Lorentz violation in a model-independent manner. Although these effective theories introduce constant background fields that carry Lorentz indices, they are not Lorentz invariant under general Lorentz transformations, but only under observer Lorentz transformations~\cite{SME-NCSM}. In general, the SME is made of pieces of the form $T^{\mu_1,\ \cdots \mu_n}{\cal O}_{\mu_1,\ \cdots \mu_n}(x)$, where the ${\cal O}_{\mu_1,\ \cdots \mu_n}(x)$ Lorentz $n$-tensors depend on the SM fields and are invariant under the SM gauge group, whereas the constant $T^{\mu_1,\ \cdots \mu_n}$ quantities transform as Lorentz $n$-tensors under observer Lorentz transformations, but they do not under the so--called particle Lorentz transformations~\cite{SME-NCSM}. The SME action can contain some $CPT$-odd terms, which necessarily implies Lorentz violation~\cite{CPTT}. The SME constitutes a valuable framework for investigating $CPT$ and Lorentz violation at the level of the SM. Investigations have been carried out in diverse scenarios, as meson systems~\cite{MSyS1,MSyS2,MSyS3,MSyS4,MSyS5}, hydrogen and antihydrogen spectroscopy~\cite{HAHE}, electromagnetic properties of the muon~\cite{MP1,MP2} and the electron~\cite{EP,CaShV}, neutrino test~\cite{NT}, electrodynamics~\cite{CFGS,KSS1,KSS2,KSS3,KSS4,CaLehPo}, the Yang--Mills theory~\cite{SanSor}, baryogenesis~\cite{BG}, study of generation of geometrical phases on the wave functions of confined electrons~\cite{GPWF}, travelling solitons~\cite{TS},  and radiative corrections~\cite{RC11,RC13,RC2,RC3,RC4,RC5,RC6,RC7}. Though the minimal SME is constructed by adding to the SM Lagrangian new observer Lorentz invariant objects of the form described above, which are renormalizable in the Dyson's sense, it can be enlarged to include nonrenormalizable interactions~\cite{IT1,IT2,BB1,BB2,BB3,CFMS1,CFMS2,Schreck1,Schreck2}.

Within the SME, the  $Z\to \gamma \gamma$ and $Z\to gg$ decays cannot be induced at the tree level, but they first emerge at the one-loop order. This is in contrast with the nonrenormalizable version of the SME or the noncommutative standard model\footnote{The differences between the nonrenormalizable version of the Standard Model Extension and the Noncommutative Standard Model has been discussed in~\cite{SMEvsNCSM}, being the latter a  subset of the former.}~\cite{NCSM}, which can generate these decays at the tree level through dimension-six gauge invariant operators~\cite{NCV}. In this paper, we will focus on one of the two {\it CPT}-odd dimension-three operators that modify the fermionic propagators~\cite{SME1,SME2,KRT1}, namely $\bar{f}\gamma_5 \slashed{b}f$, with $f$ standing for a charged lepton or a quark, and $b$ a constant four-vector with units of mass. The exact fermionic propagator that results from the introduction of this anomalous term has been already calculated and used in radiative corrections~\cite{RC2,RC3,RC4}. In this paper we will follow an alternative approach by treating this $CPT$-odd effect as a perturbation~\cite{RC5}, for this class of new physics is expected to be minuscule. The purpose of this work is twofold, as, for one side, we will estimate the size of these couplings, whereas, on the other side, we will take advantage of this problem to illustrate how to carry out radiative corrections in the context of the SME, which, unlike conventional renormalizable theories, becomes a formidable task.

The rest of the paper has been organized as follows. In Sec. \ref{ZD}, a comprehensive description of the calculations involved in the one--loop $Z\to \gamma \gamma$ and $Z\to gg$ decays is presented. Sec. \ref{D} is devoted to discuss our results. Finally, in Sec. \ref{C}, the conclusions are presented.

\section{The $Z\to \gamma \gamma$ and $Z\to gg$ decays}
\label{ZD}As commented in the introduction, the $Z\to \gamma \gamma$ and $Z\to gg$ decays cannot be induced by the SME at the tree level, but they can be generated at the one-loop level via the $CPT$-odd coupling $\bar{f}\gamma_5 \slashed{b}f$, with $f$ a charged lepton or quark and $b^\alpha$ a background field. The part of the SME which will be needed to calculate these decays is given by the following Lagrangian:
\begin{equation}
    {\cal L}=\bar{f}\left(i \slashed{D} +  \slashed{b}\gamma _5-m_f \right)f\, ,
\end{equation}
where $\slashed{D}=\gamma^\mu D_\mu$, with
\begin{equation}
D_\mu=\partial_\mu - ieQ_fA_\mu - {ig\over 2c_W}Z_\mu \left( g_V^f - g_A^f \gamma_5 \right) -ig_s{\lambda^a \over 2}G_\mu^a \, .
\end{equation}
In this expression, $A_\mu$, $Z_\mu$, and $G^a_\mu$ are the electromagnetic field, the $Z$ gauge boson, and the gluons, respectively. In addition, $Q_f$ is the electric charge of the $f$ fermion in units of $e$ and
\begin{eqnarray}
&& g_V^f=T_3^f-2Q_f\hspace{.09cm} s_W^2\, ,\\
\label{2.3}
&& g_A^f=T_3^f\, ,
\end{eqnarray}
where $T_3^f=-{1\over 2}$ for charged leptons and quarks of type down, whereas that $T_3^f={1\over 2}$ for quarks of type up. We will use the short-hand notation $s_W(c_W)$ to denote the $\sin\theta_W (\cos\theta_W)$ of the weak angle. In addition, $g_s$ is the strong constant coupling and ${\lambda^a \over 2}$, with $\lambda^a$ the Gell-man matrices, are the generators of the $SU_C(3)$ group given in its fundamental representation.

In the SME, the presence of constant background fields introduces nontrivial modifications on propagators and free-field solutions arising from the quadratic Lagrangian. As it is discussed in Ref.~\cite{CLP}, these modifications lead to nontrivial changes in the Lehmann-Symanzik-Zimmermann (LSZ) reduction formalism. In this scenario, dispersion relations can be altered significantly by allowing the occurrence of some processes forbidden in the standard theory~\cite{CLP}. In the case at hand, the complete propagator of the fermion $f$ that results from the incorporation of the anomalous $CPT$-odd bilinear term has already been calculated and used in radiative corrections~\cite{RC2,RC3,RC4}. The use of this exact propagator may induce the decay of the $Z$ boson into a pair of top quarks, which would show up through an imaginary contribution to the loop amplitudes characterizing the $Z\to \gamma \gamma$ and $Z\to gg$ transitions. Of course, a treatment of these decays using Lorentz asymmetric propagators would be interesting, but this lies outside our present scope. In this paper, we shall adopt a perturbative approach, which consist in assuming standard propagators and free-field solutions. The bilinear $CPT$-violating effect is treated by inserting the $i\lambda \slashed{b}\gamma _5$ vertex in each fermionic line~\cite{RC5} of triangle diagrams that contribute to the $Z\to \gamma \gamma$ and $Z\to gg$ decays, as it is shown in Fig.~\ref{DD}. Here, $\lambda$ is a real parameter that is introduced to our calculation for control purposes. We assume that the background field $b$ is small, and we thus investigate its effects as a first order perturbation through the aforementioned insertions.

\subsection{General structure of the $Z\gamma \gamma$ and $Zgg$ couplings}
At one--loop, the structure of the $Z\gamma \gamma$ and $Zgg$ couplings would be dictated by Bose statistics, observer Lorentz invariance, and gauge invariance. At first order in $b$, these one--loop couplings must be determined by appropriate products of the covariant objets $b_\alpha$, $Z_\alpha$, $F_{\mu \nu}$ and $\tilde{F}_{\mu \nu}=(1/2)\epsilon_{\mu \nu \lambda \rho}F^{\lambda \rho}$, in the case of the $Z\gamma \gamma$ coupling, and  $G_{\mu \nu}$ and $\tilde{G}_{\mu \nu}=(1/2)\epsilon_{\mu \nu \lambda \rho}G^{\lambda \rho}$, with $G_{\mu \nu}=T^a G^a_{\mu\nu}$, in the case of the $Zgg$ vertex. The Lorentz structure of these couplings depends crucially on the vector, $\gamma_\mu$, and axial-vector, $\gamma_\mu\gamma_5$, couplings of the $Z$ gauge boson to pairs of fermions. In the case of the vectorial contribution, we note that it is proportional to Dirac's traces involving a $\gamma_5$, so this contribution must be proportional to the Levi-Civita tensor $\epsilon_{\mu \nu \lambda \rho}$. Consequently, the interactions characterizing this amplitude must involve the dual tensor  $\tilde{F}_{\mu \nu}$ or $\tilde{G}_{\mu \nu}$. It is not difficult to convince ourselves that, at the lowest dimension, there are only two independent operators satisfying all the above requirements. In the case of the $Z\gamma \gamma$ coupling, we have
 \begin{eqnarray}
&&\tilde{\mathcal{O}}_1=b_\sigma Z^\sigma \tilde{F}_{\lambda\rho}F^{\lambda\rho}\, , \label{O1}\\
&&\tilde{\mathcal{O}}_2=b_\sigma Z^\rho \tilde{F}_{\lambda\rho}F^{\lambda\sigma}\, , \label{O2}
\end{eqnarray}
which are of dimension five (or six, if it is taken into account that $b$ has dimension one). As far as the $Zgg$ vertex is concerned, it is characterized by operators identical to the ones given above, but with the following replacements: $\tilde{F}_{\lambda\rho}F^{\lambda\rho}\to {\rm Tr}[\tilde{G}_{\lambda\rho}G^{\lambda\rho}]$, etc., where ``Tr'' denotes trace on the $SU_C(3)$ group. From these interactions, one can derive the following gauge structures on the momentum space:
\begin{eqnarray}
&&\tilde{P}_{1\,\alpha\mu\nu}={1\over m_Z^3}b_\alpha \epsilon_{\mu\nu\lambda\rho}k_1^\lambda k_2^\rho\,  ,\\
&&\tilde{P}_{2\,\alpha\mu\nu}={1\over m_Z^3}\left( b_\nu \epsilon_{\alpha\mu\lambda\rho}k_1^\lambda k_2^\rho - b_\mu \epsilon_{\alpha\nu\lambda\rho}k_1^\lambda k_2^\rho +b\cdot k_2 \epsilon_{\alpha\mu\nu\lambda}k_1^\lambda - b\cdot k_1 \epsilon_{\alpha\mu\nu\lambda}k_2^\lambda \right)\, ,
\end{eqnarray}
which have been conveniently normalized using the $m_Z$ mass. These Lorentz tensors are gauge structures in the sense that they satisfy the following Ward identities:
\begin{eqnarray}
k_1^\mu \tilde{P}_{i\,\alpha\mu\nu}&=&0 \, , \label{W1}\\
k_2^\nu \tilde{P}_{i\,\alpha\mu\nu}&=&0\, ,\label{W2}
\end{eqnarray}
for $i=1,2$. As far as the axial-vector contribution is concerned, the corresponding $Z\gamma\gamma$ and $Zgg$ amplitudes are induced by a Dirac's trace which does not involve the $\gamma_5$ matrix, so that the tensors $\tilde{F}_{\mu \nu}$ or $\tilde{G}_{\mu \nu}$ cannot emerge.
In this case, the  dimension--five operators that are analogous to those given in Eqs.~(\ref{O1},\ref{O2}) are
\begin{eqnarray}
&&\mathcal{O}_1=b_\sigma Z^\sigma F_{\lambda\rho}F^{\lambda\rho}\, ,\\
&&\mathcal{O}_2=b_\sigma Z^\rho F_{\lambda\rho}F^{\lambda\sigma}\, ,
\end{eqnarray}
and similar expressions with $G_{\mu \nu}$ instead of $F_{\mu \nu}$ for the $Zgg$ coupling. In this case, the corresponding gauge structures are given by
\begin{eqnarray}
P_{1\,\alpha\mu\nu}&=&{1\over m_Z^3}b_\alpha \left(k_{2\,\mu}k_{1\,\nu}-k_1\cdot k_2\; g_{\mu\nu} \right)\, ,\\
P_{2\,\alpha\mu\nu}&=&{1\over m_Z^3}\Big[ b_\mu \left(k_1\cdot k_2\; g_{\alpha\nu}+k_{1\,\alpha}k_{1\,\nu} \right) + b_\nu \left(k_1\cdot k_2\; g_{\alpha\mu}+k_{2\,\alpha}k_{2\,\mu} \right)\\
&&-b\cdot k_1 \left(k_{2\,\mu}g_{\alpha\nu}+k_{1\,\alpha}g_{\mu\nu} \right)-b\cdot k_2 \left(k_{1\,\nu}g_{\alpha\mu}+k_{2\,\alpha}g_{\mu\nu} \right) \Big]\, ,
\end{eqnarray}
which also satisfy Ward identities of the type given by Eqs.~(\ref{W1},\ref{W2}).

In brief, the $Z\to \gamma \gamma$ and $Z \to gg$ decays must satisfy the following requirements: 1) since only vertices of dimension less or equal than four are considered, amplitudes free of ultraviolet divergences are expected; 2) the amplitudes must be gauge invariant, that is, they must satisfy Ward identities of the type given by Eqs. (\ref{W1},\ref{W2}); 3) the amplitudes must satisfy Bose statistics, which means that the tensors characterizing the corresponding  vertex functions  must be symmetric under the interchange $\mu\leftrightarrow \nu$ and $k_1\leftrightarrow k_2$. We display, in Table~\ref{T1}, the transformation properties of these operators under the discrete operations $C$, $P$, $T$, $CP$ and $CPT$.
\begin{table}[ht]
\tbl{Transformation properties of the operators ${\cal O}_1$, ${\cal O}_2$, $\tilde{{\cal O}}_1$, and $\tilde{{\cal O}}_2$ under the discrete operations of $C$, $P$, and $T$.}
{\begin{tabular}{| c | c | c | c | c | c | c | c | c | c | c | c | }
\hline
${\cal O}_1$, ${\cal O}_2$ & $C$ & $P$ & $T$ & $CP$ & $CPT$ & $\tilde{\cal O}_1$, $\tilde{\cal O}_2$  & $C$ & $P$ & $T$ & $CP$ & $CPT$
\\
\hline
$b^0$  & + & + & + & + & + & $b^0$  & + & $-$ & $-$ & $-$ & +
\\
\hline
$\vec{b}$ & + & $-$ & $-$ & $-$ & + & $\vec{b}$ & + & + & + & + & +
\\
\hline
\end{tabular}
\label{T1}}
\end{table}
Each operator can be split into two parts: a term involving the time component $b_0$ and a term that includes the vector components $b_j$. In this table, the column ${\cal O}_1$, ${\cal O}_2$ includes the labels $b^0$ and $\vec{b}$, which represent each of these terms of the operators. The signs located to the right of the $b^0$ and $\vec{b}$ labels indicate the transformation properties of its corresponding terms. The same explanation goes for the column $\tilde{\cal O}_1$, $\tilde{\cal O}_2$.

\subsection{ Decay $Z\to \gamma \gamma$}
\begin{figure}
\center
\includegraphics[width=4.25in]{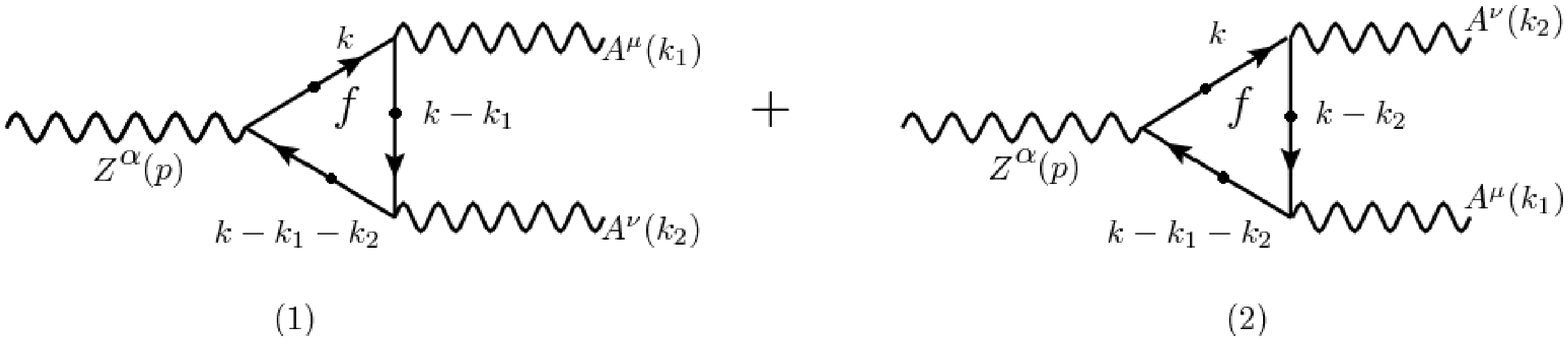}
\caption{\label{DD}Diagrams contributing to the $Z\to \gamma \gamma$ decay in the context of the SME. Dots denote insertions of the CPT-odd interaction.}
\end{figure}

The $Z\to \gamma \gamma$ decay occurs through the triangle diagrams shown in Fig.~\ref{DD}.
As it can be appreciated from this Figure, the term that introduces $CPT$ violation has been incorporated via an insertion in each fermionic line~\cite{RC5}, so the particles circulating in the loop can be described by propagators which emerge from adding to the standard propagator the insertion of the anomalous vertex:
\begin{equation}
\frac{\left(\slashed{q} +m_f\right)}{q^2-m^2_f}+\frac{\left(\slashed{q} +m_f\right)}{q^2-m^2_f}\left(i\lambda\,\slashed{b} \,\gamma_5\right)\frac{\left(\slashed{q} +m_f\right)}{q^2-m^2_f}\, ,
\end{equation}
that is, we describe the fermionic lines by effective propagators of the way
\begin{equation}
{i\Lambda (q)\over \Delta^2(q)}\, ,
\end{equation}
where
\begin{eqnarray}
\Lambda(q)&=&\left(\slashed{q} +m_f\right)\left(\slashed{q} -m_f-\lambda\,\slashed{b} \,\gamma_5\right)\left(\slashed{q} +m_f\right)\, ,\\
\Delta(q)&=&q^2-m_f^2\, ,
\end{eqnarray}
with $q$ the momentum circulating  in the loop. Notice that the standard propagator is recovered for $\lambda=0$.

The invariant amplitude for the $Z\to \gamma \gamma$ decay can be written in the way
\begin{equation}
{\cal M}={e^2g \over 2c_W}\displaystyle\sum_{f=q,l}Q_f^2 N_c  \hspace{.2cm}\Gamma_{\alpha \mu \nu}(k_1,k_2)\hspace{.2cm}\epsilon^\alpha (p,\lambda)\hspace{.2cm}\epsilon^{*\mu}(k_1,\lambda_1)\hspace{.2cm}\epsilon ^{*\nu}(k_2,\lambda_2)\, ,
\end{equation}
where $\epsilon^\alpha (p,\lambda)$, $\epsilon^{*\mu}(k_1,\lambda_1)$, and $\epsilon ^{*\nu}(k_2,\lambda_2)$ are the polarization vectors of the $Z$ gauge boson and of the pair of photons. In addition, $N_c$ is the color index, which equals $3$ for quarks and 1 for charged leptons. Also, we have included the  factor $-1$ associated with a loop of fermions. In the above expression, $\Gamma_{\alpha \mu \nu}(k_1,k_2)$ is the tensorial amplitude of the process, which is given by
\begin{equation}
\Gamma_{\alpha \mu \nu}(k_1,k_2)= \int {d^4k\over (2\pi)^4}\left({T_{\alpha \mu \nu}^{(1)}(k_1,k_2)\over \Delta^2\,\Delta^2_1\,\Delta_{12}^2}+{ T_{\alpha \mu \nu}^{(2)}(k_1,k_2)\over \Delta^2\,\Delta^2_2\,\Delta_{12}^2} \right)\, ,
\end{equation}
where
\begin{eqnarray}
T^{(1)}_{\alpha\mu\nu}={\rm Tr}\left\{\Lambda\,\gamma_\alpha\,\left(g_\nu^f-g_A^f\,\gamma_5\right)\,\Lambda_{12}\,\gamma_\nu\,\Lambda_1\,\gamma_\mu\right\}\, ,\\
T^{(2)}_{\alpha\mu\nu}={\rm Tr}\left\{\Lambda\,\gamma_\alpha\,\left(g_\nu^f-g_A^f\,\gamma_5\right)\,\Lambda_{12}\,\gamma_\mu\,\Lambda_2\,\gamma_\nu\right\}\, .
\end{eqnarray}
In addition,
\begin{eqnarray}
\Lambda&=&\left( \slashed{k}+m_f\right)\left(\slashed{k}-m_f-\lambda\,\slashed{b}\,\gamma_5\right)\left(\slashed{k}+m_f\right)\, ,
\label{N1}\\
\Lambda_1&=&\left(\slashed{k}-\slashed{k_1}+m_f\right)\left(\slashed{k}-\slashed{k_1}-m_f-\lambda\,\slashed{b}\,\gamma_5\right)\left(\slashed{k}-
\slashed{k_1}+m_f\right)\, ,\label{N2} \\
\Lambda_2&=&\left(\slashed{k}-\slashed{k_2}+m_f\right)\left(\slashed{k}-\slashed{k_2}-m_f-\lambda\,\slashed{b}\,\gamma_5\right)\left(\slashed{k}-
\slashed{k_2}+m_f\right)\, ,\label{N3}\\
\Lambda_{12}&=&\left(\slashed{k}-\slashed{k_1}-\slashed{k_2}+m_f\right)\left(\slashed{k}-\slashed{k_1}-\slashed{k_2}-m_f-\lambda\,
\slashed{b}\,\gamma_5\right)\left(\slashed{k}-\slashed{k_1}-\slashed{k_2}+m_f\right)\, ,\label{N4}\\
\Delta&=&k^2-m_f^2\, ,\\
\Delta_1&=&(k-k_1)^2-m_f^2\, ,\\
\Delta_2&=&(k-k_2)^2-m_f^2\, ,\\
\Delta_{12}&=&(k-k_1-k_2)^2-m_f^2\, .
\end{eqnarray}
In the above expressions, the superscripts $(1)$ and $(2)$ in traces stand for diagrams $(1)$ and $(2)$, respectively. Once performed some algebraic manipulations, the expressions in Eqs. (\ref{N1},\ref{N2},\ref{N3},\ref{N4}) can be rewritten as
\begin{eqnarray}
\Lambda&=&\left[\Delta-\,\lambda\,\left(\slashed{k}+m_f\right)\,\slashed{b}\,\gamma_5\right]\,\left(\slashed{k}+m_f\right)\, ,\\
\Lambda_1&=&\left[\Delta_1-\,\lambda\,\left(\slashed{k}-\slashed{k_1}+m_f\right)\,\slashed{b}\,\gamma_5\right]\,\left(\slashed{k}-
\slashed{k_1}+m_f\right)\, ,\\
\Lambda_2&=&\left[\Delta_2-\,\lambda\,\left(\slashed{k}-\slashed{k_2}+m_f\right)\,\slashed{b}\,\gamma_5\right]\,\left(\slashed{k}-
\slashed{k_2}+m_f\right)\, ,\\
\Lambda_{12}&=&\left[\Delta_{12}-\,\lambda\,\left(\slashed{k}-\slashed{k_1}-\slashed{k_2}+m_f\right)\,\slashed{b}\,\gamma_5\right]\,\left(
\slashed{k}-\slashed{k_1}-\slashed{k_2}+m_f\right)\, .
\end{eqnarray}
To simplify the analysis, we make the change of variable $k\to -k+k_1+k_2$ in the amplitude for the diagram $(2)$. Under this transformation, $\Delta_2\to \Delta_1$, $\Delta_{12}\to \Delta$, $\Delta \to \Delta_{12}$, so we have a common denominator in the complete amplitude, which now takes the way
\begin{equation}
\Gamma_{\alpha\mu\nu}(k_1,k_2)=\displaystyle\int{d^4k\over(2\pi)^4}\:{N_{\alpha \mu \nu}\over  \Delta^2\Delta_1^2\Delta_{12}^2}\, ,
\end{equation}
where
\begin{equation}
N_{\alpha \mu \nu}=T^{(1)}_{\alpha\mu\nu}+T^{(2)}_{\alpha\mu\nu}\,(k\rightarrow\,-k+k_1+k_2)\, .
\end{equation}
After calculating the Dirac's traces appearing in this expression, using de FeynCalc computer program~\cite{FC}, this tensor can be organized in powers of $\lambda$ or, equivalently, in powers of the vector $b$, as follows:
\begin{equation}
N_{\alpha \mu \nu}=N^{(0)}_{\alpha \mu \nu}+\lambda\, N^{(1)}_{\alpha \mu \nu}+\lambda^2 \, N^{(2)}_{\alpha \mu \nu}+
\lambda^3\, N^{(3)}_{\alpha \mu \nu}\, .
\end{equation}
In this expression, $N^{(0)}_{\alpha \mu \nu}$ represents the SM part, which does not contribute to the process due to the Landau-Yang's theorem. On the other hand, since the $CPT$-violating effects are expected to be minuscule,
we will work exclusively with the linear contributions from the $b$ field, which correspond to the first order in $\lambda$, that is, we will ignore the contributions given by the terms $ N^{(2)}_{\alpha \mu \nu}$ and  $ N^{(3)}_{\alpha \mu \nu}$. At this order, one has
\begin{equation}
\Gamma^{(1)}_{\alpha\mu\nu}=\displaystyle\int\,{d^4\,k\over(2\pi)^4}\:{N^{(1)}_{\alpha\mu\nu}\over \Delta^2\,\Delta^2_1\,\Delta^2_{12}}\, .
\label{eq10}
\end{equation}
In the $N^{(1)}_{\alpha\mu\nu}$ term the transversality conditions $p_\alpha=(k_1+k_2)_\alpha\to 0$, $k_{1\,\mu}\to 0$, $k_{2\,\nu}\to 0$, as well as the on-shell conditions $k_1^2=0$, $k^2_2=0$, $2\,  k_1\cdot k_2=m_Z^2$, have been imposed.

At first order in $b$, the amplitude can be divided into those parts coming from the vector and axial-vector couplings of the $Z$ gauge boson to fermions, as follows
\begin{equation}
\Gamma_{\alpha \mu \nu}^{(1)}=\Gamma_{\alpha \mu \nu}^V+\Gamma_{\alpha \mu \nu}^A\, ,
\end{equation}
where the vector and axial-vector parts are, respectively, given by
\begin{equation}
\label{AV}
\Gamma^V_{\alpha\mu\nu}=g_V^f \displaystyle\int\,{d^4\,k\over(2\pi)^4}\:{N^V_{\alpha\mu\nu}\over \Delta^2\,\Delta^2_1\,\Delta^2_{12}}\, ,
\end{equation}
\begin{equation}
\label{AAV}
\Gamma^A_{\alpha\mu\nu}=g_A^f \displaystyle\int\,{d^4\,k\over(2\pi)^4}\:{N^A_{\alpha\mu\nu}\over \Delta^2\,\Delta^2_1\,\Delta^2_{12}}\, .
\end{equation}

We turn now to solve the loop amplitudes given by Eqs.(\ref{AV},\ref{AAV}). The calculation of these amplitudes is certainly a challenging task, as they involve a huge amount of tensor integrals of diverse ranks. Therefore, the use of a computer code is mandatory. We have used the FeynCalc~\cite{FC} computer program in all stages of this calculation. However, since the Passarino--Veltaman covariant decomposition~\cite{PV}, codified in this program, does not work in this case due to the presence of quadratic powers of propagators in the amplitudes, we have instead generated a code in the environment of this program to implement the Feynman parameters technique. We now proceed to outline the main steps of this calculation. To this end, it is convenient to organize the diverse type of integrals appearing in $\Gamma^V_{\alpha\mu\nu}$ and $\Gamma^A_{\alpha\mu\nu}$ into four groups:\\

\noindent \textbf{Group I}. These integrals are of the way

\begin{equation}
 I_{\alpha\mu\nu}^\mathrm{I} \equiv  \displaystyle\int\,{d^4\,k\over(2\pi)^4}\:\left({\mathcal{O}^1_{\alpha\mu\nu}\over \Delta \Delta_1}+{\mathcal{O}^2_{\alpha\mu\nu}\over \Delta^2 \Delta_1}+{\mathcal{O}^3_{\alpha\mu\nu}\over \Delta \Delta_1^2}\right)\, .
\end{equation}
After using a convenient Feynman's parametrization, one has
\begin{equation}
I_{\alpha\mu\nu}^\mathrm{I}=\Gamma(3)  \displaystyle\int_0^1 dx\; \displaystyle\int\,{d^D\,k\over(2\pi)^4}\:{N_{\alpha\mu\nu}^I\over [(k-l_1)^2-\mathbb{R}_1]^3}\, ,
\end{equation}
where
\begin{equation}
N_{\alpha\mu\nu}^\mathrm{I} \equiv {\Gamma(2)\over \Gamma(3)}\,[(k-l_1)^2-\mathbb{R}_1]\,\mathcal{O}^1_{\alpha\mu\nu}+(1-x)\,\mathcal{O}^2_{\alpha\mu\nu}+x\,\mathcal{O}^3_{\alpha\mu\nu}\, .
\end{equation}
with $l_1=(1-x)k_1$  and $\mathbb{R}_1=m_f^2$. Notice that we have extended the measure of the momentum space integral to $D$ dimensions because the integral $ I_{\alpha\mu\nu}^\mathrm{I}$ has divergences. Of course, these divergences will be canceled by divergent terms appearing in other groups of integrals that will be listed below. \\

\noindent \textbf{Group II}. In this group there appear integrals of the way
\begin{equation}
 I_{\alpha\mu\nu}^\mathrm{II} \equiv  \displaystyle\int\,{d^4\,k\over(2\pi)^4}\:\left({\mathcal{O}^4_{\alpha\mu\nu}\over \Delta \Delta_{12}}+{\mathcal{O}^5_{\alpha\mu\nu}\over \Delta^2 \Delta_{12}}+{\mathcal{O}^6_{\alpha\mu\nu}\over \Delta \Delta_{12}^2}\right)\, ,
\end{equation}
which, after introducing a Feynman's parametrization, takes the way
\begin{equation}
I_{\alpha\mu\nu}^\mathrm{II}=\Gamma(3)  \displaystyle\int_0^1 dx\; \displaystyle\int\,{d^D\,k\over(2\pi)^4}\:{N_{\alpha\mu\nu}^{II}\over [(k-l_2)^2-\mathbb{R}_2]^3}\, ,
\end{equation}
where
\begin{equation}
N_{\alpha\mu\nu}^\mathrm{II} \equiv {\Gamma(2)\over \Gamma(3)}\,[(k-l_2)^2-\mathbb{R}_2]\,\mathcal{O}^4_{\alpha\mu\nu}+(1-x)\,\mathcal{O}^5_{\alpha\mu\nu}+x\,\mathcal{O}^6_{\alpha\mu\nu}\, ,
\end{equation}
with
\begin{eqnarray}
&&l_2=(1-x)(k_1+k_2)\, ,\\
&&\mathbb{R}_2=m_f^2\,\left[1-{4\over \tau_f}\,x\,(1-x)\right]\, .
\end{eqnarray}
In addition, $\tau_f\equiv {4m_f^2\over m_Z^2}$.\\

\noindent \textbf{Group III}. The integrals of this classification have the form

\begin{equation}
 I_{\alpha\mu\nu}^\mathrm{III} \equiv  \displaystyle\int\,{d^4\,k\over(2\pi)^4}\:\left({\mathcal{O}^7_{\alpha\mu\nu}\over \Delta_1 \Delta_{12}}+{\mathcal{O}^8_{\alpha\mu\nu}\over \Delta^2_1 \Delta_{12}}+{\mathcal{O}^9_{\alpha\mu\nu}\over \Delta_1 \Delta_{12}^2}\right)\, ,
\end{equation}
which, after introducing a Feynman's parametrization, takes the way
\begin{equation}
I_{\alpha\mu\nu}^\mathrm{III}=\Gamma(3)  \displaystyle\int_0^1 dx\; \displaystyle\int\,{d^D\,k\over(2\pi)^4}\:{N_{\alpha\mu\nu}^{III}\over [(k-l_3)^2-\mathbb{R}_3]^3}\, ,
\end{equation}
where
\begin{equation}
N_{\alpha\mu\nu}^\mathrm{III} \equiv {\Gamma(2)\over \Gamma(3)}\,[(k-l_3)^2-\mathbb{R}_3]\,\mathcal{O}^7_{\alpha\mu\nu}+(1-x)\,\mathcal{O}^8_{\alpha\mu\nu}+x\,\mathcal{O}^9_{\alpha\mu\nu}\, ,
\end{equation}
with $l_3=k_1+(1-x)k_2$ and $\mathbb{R}_3=m_f^2$.\\

\noindent \textbf{Group IV}. This group has integrals that are expressed as
\begin{equation}
 I_{\alpha\mu\nu}^\mathrm{IV} \equiv  \displaystyle\int\,{d^4\,k\over(2\pi)^4}\:\left({\mathcal{O}^{10}_{\alpha\mu\nu}\over \Delta \Delta_1 \Delta_{12}}+{\mathcal{O}^{11}_{\alpha\mu\nu}\over \Delta^2 \Delta_1 \Delta_{12}}+{\mathcal{O}^{12}_{\alpha\mu\nu}\over \Delta \Delta^2_1 \Delta_{12}}+{\mathcal{O}^{13}_{\alpha\mu\nu}\over \Delta \Delta_1 \Delta^2_{12}}\right)\, .
\end{equation}
After performing a Feynman's parametrization, one has
\begin{equation}
I_{\alpha\mu\nu}^\mathrm{IV}=\Gamma(4)  \displaystyle\int_0^1 dx\; \int_0^{1-x} dy\;\displaystyle\int\,{d^D\,k\over(2\pi)^4}\:{N_{\alpha\mu\nu}^{IV}\over [(k-l)^2-\mathbb{R}]^4}\, ,
\end{equation}
where
\begin{equation}
N_{\alpha\mu\nu}^\mathrm{IV} \equiv \,[(k-l)^2-\mathbb{R}]\,\mathcal{O}^{10}_{\alpha \mu \nu}+(1-x-y)\,\mathcal{O}^{11}_{\alpha \mu \nu}+x\,\mathcal{O}^{12}_{\alpha \mu \nu}+y\,\mathcal{O}^{13}_{\alpha \mu \nu}\, ,
\end{equation}
with
\begin{eqnarray}
&& l=(x+y)k_1+yk_2\, ,\\
&&\mathbb{R}=m_f^2 \,\left[1-{4\over \tau_f}\,y\,(1-x-y)\right]\, .
\end{eqnarray}

Adding together all the above types of integrals, one has
\begin{equation}
\Gamma^A_{\alpha \mu \nu }=g_A^f\left(I^{I}_{\alpha \mu \nu}+I^{II}_{\alpha \mu \nu}+I^{III}_{\alpha \mu \nu}+
I^{IV}_{\alpha \mu \nu}\right)\, .
\end{equation}
The structure of the amplitude $\Gamma^V_{\alpha \mu \nu}$ is identical to that of $\Gamma^A_{\alpha \mu \nu}$ but with the $\tilde{\mathcal{O}}^i_{\alpha \mu \nu}$ tensors instead of the $\mathcal{O}^i_{\alpha \mu \nu}$ ones. Although it is not evident, there is no contribution to the $\Gamma^V_{\alpha \mu \nu}$ amplitude. To show this, we used the relations given in \ref{A}, together with the Schouten's identity, which is given by
\begin{equation}
g_{\alpha\beta}\epsilon_{\mu\nu\lambda\rho}+g_{\alpha\lambda}\epsilon_{\rho\beta\mu\nu}+
g_{\alpha\mu}\epsilon_{\nu\lambda\rho\beta}+g_{\alpha\nu}\epsilon_{\lambda\rho\beta\mu}+g_{\alpha\rho}\epsilon_{\beta\mu\nu\lambda}=0.
\end{equation}
The fact that $\Gamma^V_{\alpha \mu \nu}=0$ means that the coupling of $Z$ to fermions that is proportional to $\gamma_\mu$ does not contribute to the $Z\to \gamma \gamma$ decay (and neither to the $Z\to gg$ decay). This in turn implies that the $\gamma^* \gamma \gamma$ coupling, with a virtual photon, is not induced at this level, which is in agrement with Furry's theorem and previous results on one-loop renormalization of Lorentz-violating electrodynamics~\cite{QEDR}.

 Returning to the $\Gamma^A_{\alpha \mu \nu }$ amplitude, it is found that the $I^{I}_{\alpha \mu \nu }$ and $I^{II}_{\alpha \mu \nu }$ integrals are separately identical to zero. On the other hand, the integral $I^{III}_{\alpha \mu \nu }$ leads to a result which is free of divergences but not gauge invariant. It is given by
\begin{eqnarray}
I^{III}_{\alpha \mu \nu }=-\frac{i}{(4\pi)^2}\,{16\,\tau_f\over \tau_f-1}\,f(\tau_f)\,b_\alpha\,g_{\mu\nu}\, ,
\end{eqnarray}
where
\begin{equation}
f(\tau_f)=\left\{
\begin{array}{clc}
\sqrt{\tau_f-1}\,\arctan \left({1\over\sqrt{\tau_f-1}}\right), && \tau_f>1\\
{1\over 2}\sqrt{1-\tau_f}\left[{\rm log}\left({1+\sqrt{1-\tau_f}\over 1-\sqrt{1-\tau_f} }\right)-i\,\pi\right], && \tau_f<1
\end{array}
\right. \, .
\end{equation}
Note that the divergence induced by the $\mathcal{O}^7_{\alpha\mu\nu}$ term is exactly canceled by the sum of divergences generated by the $\mathcal{O}^8_{\alpha\mu\nu}$ and $\mathcal{O}^9_{\alpha\mu\nu}$ terms. As far as the $I^{IV}_{\alpha \mu \nu }$ integral is concerned, it leads to a result which is free of divergences but not gauge invariant. However, when this result is added to the one given by $I^{III}_{\alpha \mu \nu }$, a gauge invariant result is obtained, which can be written as
\begin{eqnarray}
\Gamma^A_{\alpha \mu \nu }=-{32\,i\,g^f_A\,m_Z\over(4\pi)^2}\,\left(\mathcal{A}_1 P_{1\alpha\mu\nu}+\mathcal{A}_2 P_{2\alpha\mu\nu}\right)\, ,
\end{eqnarray}
where
\begin{eqnarray}
\mathcal{A}_1&=&{f(\tau_f)\over \tau_f-1}\left\{\tau_f\,[1-f(\tau_f)]-1\right\}\, , \nonumber \\
\mathcal{A}_2 &=&1-f(\tau_f)\, .
\end{eqnarray}
As expected, our result is free of ultraviolet divergences. The one-loop divergent structure of the SME has already been studied in its QED~\cite{QEDR}, electroweak~\cite{EWR}, and QCD~\cite{QCDR} sectors. On the other hand, the presence of a mass independent term in the ${\cal A}_2$ amplitude constitutes the well-known ABJ anomaly. In our case, the anomaly vanishes if the background field $b_\alpha$ is assumed to be the same for each class of family (of leptons and quarks) since
\begin{equation}
\sum_{f=l_i,u_i,d_i}g^f_AQ^2_fN_C=0\, , \ \ \ \ \ i=1,2,3 \, .
\end{equation}
Otherwise, there must exist a correlation among the different $b$'s of each type of family.

Then, the invariant amplitude for the  $Z\rightarrow\gamma\gamma$ decay can be written as

\begin{equation}
{\cal M}=-{16i\, \alpha^{3/2}m_Z\over s_{2W}\sqrt{\pi}} \sum_{f=l,q}\left({\cal F}^f_1 P^f_{1\, \alpha \mu \nu}+{\cal F}^f_2 P^f_{2\, \alpha \mu \nu}\right)\epsilon^\alpha(p,\,\lambda)\epsilon^{\mu*}(k_1,\,\lambda_1)\epsilon^{\nu*}(k_2,\,\lambda_2)\, ,
\end{equation}
where
\begin{equation}
{\cal F}^f_i \equiv g^f_A Q^2_f N_c {\cal A}_i\, , \, \, \, \, \, i=1,2 \, .
\end{equation}
We have introduced the superscript $f$ in the gauge structures $P^f_{i\, \alpha \mu \nu}$ to emphasize that, in general, there is a distinct four-vector $b^f_\alpha$ for each fermionic flavor.

Squaring the invariant amplitude yields
\begin{eqnarray}
|\overline{\cal M}|^2&=&\left({1\over 3}\displaystyle\sum_{\lambda=1}^3\right)\left(\displaystyle\sum_{\lambda_1=1}^2\right)\left(\displaystyle\sum_{\lambda_2=1}^2\right){\cal M}{\cal M}^\dagger \nonumber \\
&=&\left({16\over s_{2W}}\right)^2{\alpha^3\,m_Z^2\over 3\pi}\sum_{f=l,q}\, \sum_{f'=l,q}\, \Big\{-\frac{1}{2}\Big[\left({\cal F}^f_1-{\cal F}^f_2\right)\left({\cal F}^{f'*}_1-{\cal F}^{f'*}_2\right)
\nonumber \\ &&
+
2{\cal F}^f_2 {\cal F}^{f'*}_2\Big]\left(\frac{b^f\cdot b^{f'}}{m^2_Z}\right)\nonumber \\
&&+\frac{1}{2}\left[{\cal F}^f_1\left({\cal F}^{f'*}_1+{\cal F}^{f'*}_2\right)+{\cal F}^f_2\left({\cal F}^{f'*}_1-{\cal F}^{f'*}_2\right)\right]
\left(\frac{b^f\cdot p\,\, b^{f'}\cdot p}{m^4_Z}\right)\nonumber \\
&&+\left({\cal F}^f_2 {\cal F}^{f'*}_2-{\cal F}^f_2{\cal F}^{f'*}_1-{\cal F}^f_1{\cal F}^{f'*}_2\right)\left(\frac{b^f\cdot k_1\, \, b^{f'}\cdot k_1 +b^f\cdot k_2\, \, b^{f'}\cdot k_2}{m^4_Z}\right)\nonumber \\
&&-\Big[{\cal F}^f_2\left({\cal F}^{f'*}_1-{\cal F}^{f'*}_2\right)+{\cal F}^{f'*}_2\left({\cal F}^{f}_1-{\cal F}^{f}_2\right)
\nonumber \\ &&
-{\cal F}^f_2{\cal F}^{f'*}_2\Big]\left(\frac{b^f\cdot k_1\, \, b^{f'}\cdot k_2 +b^f\cdot k_2\, \, b^{f'}\cdot k_1}{m^4_Z}\right)
  \Big\}\, .
\end{eqnarray}
This result is valid in any inertial frame. It is interesting to discuss this result from the perspective of observer Lorentz transformations. Under this type of transformations, the constant background field $b^f$ transforms as four-vector, although under the particle Lorentz transformations the four components of $b^f_\alpha$ transform as scalars~\cite{SME1,SME2}. Since the SME is invariant under observer Lorentz transformations (observations in two inertial frames are connected through coordinate changes), it is possible to analyze the above squared amplitude in some particular reference frames. For instance, it is interesting to analyze this result from the perspective of time-like or space-like background $b$ fields. The case of light-like background $b$ fields is interesting and will be analyzed too. To carry out this analysis for the cases of time-like or space-time background fields, we need to assume the presence of an unique $b$. Then, assume that $b$ is time-like, that is, $b^2>0$. In this case, we can pass to a new reference frame through a boost with vector parameter $\mathbf{\beta}=\mathbf{b}/b^0$ in which $b'^\alpha=(b',0,0,0)$. Notice that in this scenario the little group of $b$ is the rotations group $SO(3)$. In this frame, the squared of the amplitude takes the way
\begin{eqnarray}
|\overline{\cal M}|^2&=&\left({16\over s_{2W}}\right)^2{\alpha^3\,m_Z^2\over 3\pi}\sum_{f=l,q}\, \sum_{f'=l,q}\, \Big[{\cal F}^f_1\left(F^{f'*}_1-F^{f'*}_2\right)\nonumber \\
&&+{\cal F}^f_2\left(3F^{f'*}_2-F^{f'*}_1\right)\Big]\, \left(\frac{b'^2_0\, \mathbf{p}'^2}{2m^4_Z}\right) \, ,
\end{eqnarray}
where $\mathbf{p}'$ is the spatial component of the momentum of the $Z$ gauge boson in this reference frame. Of course, $\mathbf{p}'=0$ is an allowed value for the spatial momentum, so the amplitude vanishes only in this special case. On the other hand, in the case of a space-like background field, $b^2<0$, it always possible to perform a boost to a new reference frame in which $b'^\alpha=(0,\mathbf{b}')$. The boost parameter is given in this case by $\mathbf{\beta}=(b^0/|\mathbf{b}|)\hat{\mathbf{b}}$, wit $\hat{\mathbf{b}}$ the unit vector in the $\mathbf{b}$ direction. Note that, in this case, the little group of $b$ is $SO(1,2)$. In this frame, the squared amplitude takes the way
\begin{eqnarray}
|\overline{\cal M}|^2&=&\left({16\over s_{2W}}\right)^2{\alpha^3\,m_Z^2\over 3\pi}\sum_{f=l,q}\, \sum_{f'=l,q}\Big\{2{\cal F}^f_2F^{f'*}_2\left(\frac{4\left(\mathbf{b}'\cdot \mathbf{k}'_1\right)\left(\mathbf{b}'\cdot \mathbf{k}'_2\right)
-m^2_Z\, \mathbf{b}'^2}{2\, m^4_Z}\right)\nonumber \\
&&+\left({\cal F}^f_1-{\cal F}^f_2\right)\left(F^{f'*}_1-F^{f'*}_2\right)\left(\frac{\left(\mathbf{b}'\cdot \mathbf{p}'\right)^2-m^2_Z\, \mathbf{b}'^2}{2\, m^4_Z}\right)
\Big\}\, .
\end{eqnarray}
From this expression, it can be seen that  there is no possible kinematical configuration that, together with an appropriate $\mathbf{b}'$ orientation, leads to a zero result. Finally, if $b^2=0$, the general expression for the squared amplitude can be written as:
\begin{eqnarray}
|\overline{\cal M}|^2&=&\left({16\over s_{2W}}\right)^2{\alpha^3\,m_Z^2\over 3\pi}\sum_{f=l,q}\, \sum_{f'=l,q}\left(\frac{b^2_0}{2\, m^2_Z}\right)\nonumber \\
&&\times \Bigg\{ \left({\cal F}^f_1-{\cal F}^f_2\right)\left({\cal F}^{f'*}_1-{\cal F}^{f'*}_2\right)\left(\frac{p^0-\hat{\mathbf{b}}\cdot \mathbf{p}}{m_Z}\right)^2\nonumber \\
&&+8{\cal F}^f_2{\cal F}^{f'*}_2\left(\frac{k^0_1-\hat{\mathbf{b}}\cdot \mathbf{k}_1}{m_Z}\right)\left(\frac{k^0_2-\hat{\mathbf{b}}\cdot \mathbf{k}_2}{m_Z}\right)
\Bigg \}\, ,
\end{eqnarray}
where, as in the previous cases, we have assumed the existence of a sole background field $b$. It is easy to see that there is no kinematic scenario in which this expression to be zero.

We now proceed to analyze our results in the $Z$ rest frame. In this frame, the squared of the amplitude is greatly simplified. In particular, the result does not depend on the temporal part of the $b^f_\alpha$ four-vector. In fact, once implemented the corresponding kinematics, one obtains
\begin{eqnarray}
|\overline{\cal M}|^2&=&\left({16\over s_{2W}}\right)^2{\alpha^3\,m_Z^2\over 3\pi}\sum_{f=l,q}\, \sum_{f'=l,q}\Big\{-\frac{1}{2}\Big[\left({\cal F}^f_1-{\cal F}^f_2\right)\left({\cal F}^{f'*}_1-{\cal F}^{f'*}_2\right)
\nonumber \\ &&
+2{\cal F}^f_2 {\cal F}^{f'*}_2\Big]\left(\frac{\mathbf{b}^f\cdot \mathbf{b}^{f'}}{m^2_Z}\right)-4{\cal F}^f_2{\cal F}^{f'}_2\left(\frac{\mathbf{b}^f\cdot \mathbf{k}_1 \, \, \mathbf{b}^{f'}\cdot \mathbf{k}_1}{m^4_Z}\right)\Big\}\, .
\end{eqnarray}
Further simplifications are obtained assuming that the vectors $\mathbf{b}^f$ are mutually orthogonal, that is,
\begin{equation}
\mathbf{b}^f\cdot \mathbf{b}^{f'}=|\mathbf{b}^f||\mathbf{b}^{f'}|\delta^{ff'}\, .
\end{equation}
If, in addition, we denote by $\theta_f$ the angle between the vectors $\mathbf{k}_1$ and $\mathbf{b}^f$, the squared amplitude takes the way
\begin{eqnarray}
|\overline{\cal M}|^2&=&\left({4\over s_{2W}}\right)^2{\alpha^3\,m_Z^2\over 3\pi}\left[\sum_{f}\Bigg[ |{\cal F}^f_1-{\cal F}^f_2|^2+2|{\cal F}^f_2|^2\sin^2\theta_f\right]\left(\frac{|\mathbf{b}^f|^2}{m^2_Z}\right)
\nonumber \\ &&
+\sum_{f\neq f'}{\cal F}^f_2{\cal F}^{f'}_2 \left(\frac{|\mathbf{b}^f||\mathbf{b}^{f'}|}{m^2_Z}\right)\sin2\theta_f
\Bigg]\, .
\end{eqnarray}
The corresponding branching ratio is given by
\begin{eqnarray}
BR(Z\to \gamma \gamma)=\frac{\alpha^3}{3\pi^2 s^2_{2W}}\left(\frac{m_Z}{\Gamma_Z}\right)\Bigg[&+&\sum_{f}\left[ |{\cal F}^f_1-{\cal F}^f_2|^2+2|{\cal F}^f_2|^2\sin^2\theta_f\right]\left(\frac{|\mathbf{b}^f|^2}{m^2_Z}\right)\nonumber \\
&+&\sum_{f\neq f'}{\cal F}^f_2{\cal F}^{f'}_2 \left(\frac{|\mathbf{b}^f||\mathbf{b}^{f'}|}{m^2_Z}\right)\sin2\theta_f
\Bigg]\, .
\end{eqnarray}
where $\Gamma_Z$ is the $Z$ total width decay.

\subsection{The $Z\to gg$ decay}
The calculation of the $Z\to gg$ amplitude is entirely similar to the one given in the case of the $Z\to \gamma \gamma$ decay, with only some slight changes. The corresponding branching ratio can be written as
\begin{eqnarray}
BR(Z\to gg)=\frac{2\alpha\, \alpha^2_s}{3\pi^2 s^2_{2W}}\left(\frac{m_Z}{\Gamma_Z}\right)\Bigg[&+&\sum_{q}\left[ |F^q_1-F^q_2|^2+2|F^q_2|^2\sin^2\theta_q\right]\left(\frac{|\mathbf{b}^q|^2}{m^2_Z}\right)\nonumber \\
&+&\sum_{q\neq q'}F^q_2F^{q'}_2 \left(\frac{|\mathbf{b}^q||\mathbf{b}^{q'}|}{m^2_Z}\right)\sin2\theta_q
\Bigg]\, ,
\end{eqnarray}
where $\alpha_s=g^2_s/4\pi$, and
\begin{equation}
F^q_i=g^q_A {\cal A}_i\, , \, \, \, \, \, i=1,2 \, .
\end{equation}

\section{Discussion}
\label{D}We now proceed to discuss our results. Previously, we have shown that in the reference frames determined by a time-like or space-like background field $b$ the squared of the amplitude does not vanish, with the exception of the kinematic value $\mathbf{p'}=0$ in the former case. Also it was shown that, in the $Z$ rest frame, the $Z\to \gamma \gamma$ and $Z\to gg$ decays only  depend on the spatial component of the four-vector $b^f_\alpha$. This result is independent of the space-time nature of the the background field $b$. Technically, the discussion can greatly be simplified if it is assumed that the two photons or two gluons are emitted along the $\vec{b}^f$ direction. In this case, the branching ratios given in previous section take a more simple way
\begin{eqnarray}
BR(Z\to \gamma \gamma)&=&\frac{\alpha^3}{3\pi^2 s^2_{2W}}\left(\frac{m_Z}{\Gamma_Z}\right)\sum_{f} \Big|{\cal F}^f_1-{\cal F}^f_2\Big|^2\left(\frac{|\mathbf{b}^f|^2}{m^2_Z}\right)\, , \\
BR(Z\to gg)&=&\frac{2\alpha\, \alpha^2_s}{3\pi^2 s^2_{2W}}\left(\frac{m_Z}{\Gamma_Z}\right)\sum_{q} \Big|F^q_1-F^q_2\Big|^2\left(\frac{|\mathbf{b}^q|^2}{m^2_Z}\right)\, .
\end{eqnarray}

In order to determine the relative importance of the contributions to these decays, arising from each type of fermion, let us analyze the behavior of the corresponding loop functions. In the heavy mass limit, one obtains
\begin{equation}
\displaystyle\lim_{\tau\rightarrow\infty}\,\left({\cal F}^f_1-{\cal F}^f_2\right)=0\, ,
\end{equation}
whereas, in the light mass limit, one has
\begin{equation}
\displaystyle\lim_{\tau\rightarrow 0}\,\left({\cal F}^f_1-{\cal F}^f_2\right)=\infty\, .
\end{equation}
These results show that the decays $Z\to \gamma \gamma$ and $Z\to gg$ are more sensitive to lighter fermions. Specifically, $|{\cal F}^e_1-{\cal F}^e_2|=24$ and $|{\cal F}^t_1-{\cal F}^t_2|=38\times 10^{-2}$ for the electron and the top quark, respectively. Using the data reported by the Particle Data Group~\cite{PDG}, we can write
\begin{eqnarray}
BR(Z\to \gamma \gamma)&=&6.7\times 10^{-6}\sum_{f} \Big|{\cal F}^f_1-{\cal F}^f_2\Big|^2\left(\frac{|\mathbf{b}^f|^2}{m^2_Z}\right)\, , \\
BR(Z\to gg)&=&3.6\times 10^{-2}\sum_{q} \Big|F^q_1-F^q_2\Big|^2\left(\frac{|\mathbf{b}^q|^2}{m^2_Z}\right)\, .
\end{eqnarray}
The background field $b^f_\alpha$ has been bounded for the case of lighter charged leptons by different means, which include Hg/Cs comparison~\cite{KRT1,Hg}, Penning trap~\cite{Pt}, torsion pendulum~\cite{tP1,tP2,tP3}, K/He magnetometer~\cite{KHe}, $g_\mu-2$ data~\cite{MMM1,MMM2,MMM3}, and mononium spectroscopy~\cite{MoS}. These constraints lie within the ranges of $|\mathbf{b}^e|<(10^{-27}-10^{-31})\, {\rm eV}$ and $|\mathbf{b}^\mu|<(10^{-22}-10^{-24})\, {\rm eV}$ for the electron and muon, respectively. The corresponding contribution to $BR(Z\to \gamma \gamma)$ are of order of $10^{-36}$ and $10^{-31}$ at most. On the other hand, one could be tempted to think that a more favorable scenario can arise from the top quark contribution, as $\mathbf{b}^t$ is much less sensitive to low-energy experiments. In this case, one has
\begin{eqnarray}
BR(Z\to \gamma \gamma)&=&1.6\times 10^{-7}\left(\frac{|\mathbf{b}^t|^2}{m^2_Z}\right)\, , \\
BR(Z\to gg)&=&4.7\times 10^{-5}\left(\frac{|\mathbf{b}^t|^2}{m^2_Z}\right)\, .
\end{eqnarray}
However, even in this case, expectations are quite poor. The reason is that naturally one can expect that $|\mathbf{b}^t|\sim m^2/M_P$, with $m$ of order of the Fermi scale, $v\simeq 246$ GeV, and $M_P$ the Planck mass, so that
\begin{eqnarray}
BR(Z\to \gamma \gamma)&\sim & 10^{-7}\left(\frac{v}{M_P}\right)^2\, , \\
BR(Z\to gg)&\sim &10^{-4}\left(\frac{v}{M_P}\right)^2\, .
\end{eqnarray}
These results reinforce the idea that $CPT$ and Lorentz violating effects would be so tiny that they could be detectable only by experiments of exceptional sensitivity.

\section{Conclusions}
\label{C} The decays $Z\to \gamma \gamma$ and $Z\to gg$ are strictly forbidden in the standard model, but they can arise in the presence of constant background fields. In this paper, these decays have been studied in the context of the renormalizable version of the standard model extension, which is an effective field theory that incorporates $CPT$-odd and Lorentz--violating effects by introducing background constant tensor fields. A background field $b_\alpha$ was considered through the bilinear $\bar{f}\gamma_5 \slashed{b}f$ interaction. It was shown that this interaction generates the $Z\to \gamma \gamma$ and $Z\to gg$ decays at the one-loop level, and that the corresponding amplitudes are gauge invariant and free of ultraviolet divergences. Since the theory is invariant under observer Lorentz transformations, it was possible to analyze the squared of the amplitude of the $Z\gamma \gamma$ and $Zgg$ interactions in the light of the time-like, space-like or light-like nature of the four-vector $b$. In the reference frame defined by a time-like background field $b$, it was found that the squared of the amplitude is proportional to $\mathbf{p'}^2$, with $\mathbf{p'}$ the spatial momentum of the $Z$ gauge boson in this frame. So, in this case, these couplings disappear only in the particular kinematic case $\mathbf{p'}=0$. On the other hand, in the frame determined by a space-like vector $b$, it was shown that these couplings do not vanish for any kinematic configuration. In the case of a light-like background field $b$, it was shown that the squared of the amplitude does not vanish in any allowed kinematic configuration. It was found that in the $Z$ rest frame, the decay widths of these decays do not depend on the time component of the background field $b_\alpha$, but only on its spatial part, being the decay widths proportional to $\mathbf{b}^2/m^2_Z$. From the point of view of  observer Lorentz transformations, $\mathbf{b}$ transform as a vector under the rotations group $SO(3)$, so that if the $Z\to \gamma \gamma$ and $ Z\to gg$ decays are considered jointly with this vector, invariance under rotations is preserved, and, from this perspective, the Landau-Yang's theorem is not violated indeed. However, from the point of view of particle Lorentz transformations, $\mathbf{b}$ does not transform, so that the existence of these decays can be seen as a violation of the Landau-Yang's theorem. Although nonzero, the branching ratios for these decays would be undetectable as it is expected that $|\mathbf{b}|$ to be extremely small.

\section*{Acknowledgments}
We acknowledge financial support from CONACYT (M\' exico). H.N.S, J.J.T., and E.S.T. also acknowledge financial support from SNI (M\' exico).

\appendix

\section{Momentum-space integrals}
\label{A} In this appendix, we present a list of the space-momentum integrals used throughout the paper.
\begin{eqnarray}
&&\hspace{-1cm}(\mu^2)^{4-D\over 2}\displaystyle\int\,{d^D\,k\over(2\pi)^D}\;{1\over (k^2-\mathbb{R})^N}
\nonumber \\ &&
={i(-1)^N\over (4\pi)^2}\,(4\pi \mu^2)^{4-D\over 2}\,{\Gamma(N-{D\over 2})\over \Gamma(N)}\left({1\over \mathbb{R}}\right)^{N-{D\over 2}}\, ,\\ \nonumber
 && \\
&&\hspace{-1cm}(\mu^2)^{4-D\over 2}\displaystyle\int\,{d^D\,k\over(2\pi)^D}\;{k^2\over (k^2-\mathbb{R})^N}
\nonumber \\ &&
={i(-1)^{N-1}\over (4\pi)^2}\,(4\pi \mu^2)^{4-D\over 2}\,{D\over 2}\,{\Gamma(N-{D\over 2}-1)\over \Gamma(N)}\left({1\over \mathbb{R}}\right)^{N-{D\over 2}}\, ,\\ \nonumber
&& \\ &&
\hspace{-1cm}(\mu^2)^{4-D\over 2}\displaystyle\int\,{d^D\,k\over(2\pi)^D}\;{k^4\over (k^2-\mathbb{R})^N}
\nonumber \\ &&
={i(-1)^{N}\over (4\pi)^2}\,(4\pi \mu^2)^{4-D\over 2}\,{D(D+2)\over 4}\,{\Gamma(N-{D\over 2}-2)\over \Gamma(N)}\left({1\over \mathbb{R}}\right)^{N-{D\over 2}}\, ,
\end{eqnarray}
where $\mu$ is the mass scale of dimensional regularization. To reduce tensor integrals to the ones of the above type, we used systematically the following relations
\begin{eqnarray}
&&k_\mu k_\nu \to {k^2\over D} g_{\mu\nu}\, ,\\
&&k_\mu k_\nu k_\alpha k_\beta \to {k^4 \over D(D+2)}\left(g_{\mu\nu}g_{\alpha\beta}+g_{\mu\alpha}g_{\beta\nu}+g_{\mu\beta}g_{\nu\alpha}\right)\, ,\\
&&k^2 k_\mu k_\nu \to {k^4\over D}g_{\mu\nu}\, .
\end{eqnarray}

\end{document}